\documentclass[preprint,review,12pt]{elsarticle}

\usepackage{amssymb}

\usepackage{algorithm}
\usepackage{algorithmic}
\usepackage{booktabs}
\usepackage{multirow}
\usepackage[table]{xcolor}
\usepackage{adjustbox}
\newcommand{\stitle}[1]{\noindent{\bf #1}}

\newcommand{\new}[1]{\textcolor{black}{ #1}}
\usepackage{newfloat}
\usepackage{listings}
\usepackage{subfigure}
\usepackage[hyphens]{url}

\journal{Decision Support Systems}

\begin{document}

\setcounter{table}{0}

\begin{frontmatter}



\title{Auditing the Imputation Effect on Fairness of Predictive Analytics in Higher Education}



\author[inst1]{Hadis Anahideh}

\affiliation[inst1]{organization={University of Illinois at Chicago, Department of Mechanical and Industrial Engineering},
            addressline={842 W Taylor St}, 
            city={Chicago},
            postcode={60607}, 
            state={IL},
            country={USA}}

\author[inst1]{Nazanin Nezami}
\author[inst1]{Parian Haghighat}
\author[inst2]{Denisa G\`andara}

\affiliation[inst2]{organization={The University of Texas at Austin, Department of Educational Leadership and Policy},
            addressline={1912 Speedway, Stop D5000}, 
            city={Austin},
            postcode={78712}, 
            state={Texas},
            country={USA}}

\begin{abstract}
Colleges and universities use predictive analytics in a variety of ways to increase student success rates. 
Despite the potential for predictive analytics, two major barriers exist to their adoption in higher education: (a) the lack of democratization in deployment, and (b) the potential to exacerbate inequalities. Education researchers and policymakers encounter numerous challenges in deploying predictive modeling in practice. These challenges present in different steps of modeling including data preparation, model development, and evaluation. Nevertheless, each of these steps can introduce additional bias to the system if not appropriately performed. Most large-scale and nationally representative education data sets suffer from a significant number of incomplete responses from the research participants. Missing Values are the frequent latent causes behind many data analysis challenges. While many education-related studies addressed the challenges of missing data, little is known about the impact of handling missing values on the fairness of predictive outcomes in practice. \new{In addition, assessing the fairness of predictive outcomes is an essential step in the development of equitable algorithms. In particular, out-of-sample evaluation provides a less biased estimate of learning performance and fairness for real-time deployment.}

In this paper, we set out to first assess the disparities in predictive modeling outcomes for college-student success, then investigate the impact of imputation techniques on the model performance and fairness using a commonly used set of metrics. \new{We conduct a prospective evaluation to provide a less biased estimation of future performance and fairness than an evaluation of historical data. Our comprehensive analysis of a real large-scale education dataset reveals key insights on modeling disparities and how imputation techniques impact the fairness of the student-success predictive outcome under different testing scenarios. Our results indicate that imputation introduces bias if the testing set follows the historical distribution. However, if the injustice in society is addressed and consequently the upcoming batch of observations is equalized, the model would be less biased. }

\end{abstract}




\end{frontmatter}


\section{Introduction}

Predictive analytics has become an increasingly hot topic in higher education.
In particular, 
predictive analytics tools have been used 
to predict various measures of student success (e.g., course completion, retention, and degree attainment) by mapping the input set of attributes of individuals (e.g., the student’s high school GPA and demographic features) with their outcomes (e.g., college credits accumulated) \cite{ekowo2016promise}. 
Campus officials have used these predictions to guide decisions surrounding college admissions and student-support interventions, such as providing more intensive advising to certain students \cite{ekowo2016promise}. 

Despite the potential for predictive analytics, there is a critical disconnection between predictive analytics in higher education research and their accessibility of them in practice. Two major barriers to existing uses of predictive analytics in higher education that cause this disconnection are the lack of democratization in deployment and the potential to exacerbate inequalities.

First, education researchers and policymakers face many challenges in deploying predictive and statistical techniques in practice. These challenges present in different steps of modeling including data cleaning (e.g. imputation), identifying the most important attributes associated with success, selecting the correct predictive modeling technique, and calibrating the hyperparameters of the selected model.
Nevertheless, each of these steps can introduce additional bias to the system if not appropriately performed \cite{barocas2016big}. Missing Values are the frequent latent causes behind many data analysis challenges. Most large-scale and nationally representative education data sets suffer from a significant number of incomplete responses from the research participants. While many education-related studies addressed the challenges of missing data, \cite{missing-review1,Missing3-MI,dataprep3}, little is known about the impact of handling missing values on the fairness of predictive outcomes in practice. 
To date, few works studied the impact of data preparation on the unfairness of the predictive outcome in a limited setting \cite{valentim2019impact} or using merely a single notion of fairness metrics \cite{missing2021}.

Second, predictive models rely on historical data and have the potential to exacerbate social inequalities \cite{ekowo2016promise,kizilcec2020algorithmic}. Over the last decade, researchers realized that disregarding the consequences and especially the societal impact of algorithmic decision-making, might negatively impact individuals' lives. COMPAS, a criminal justice support tool, was found to be decidedly biased against Black people \cite{propublica}. 
Colleges and universities have been using risk algorithms to evaluate their students. Recently, Markup investigated four major public universities and found that EAB's Navigate software is racially biased \cite{markup}. 
Ensuring fair and unbiased assessment, however, is complex and it requires education researchers and practitioners to undergo a comprehensive algorithm audit to ensure the technical correctness and social accountability of their algorithms.

It is imperative that predictive models are designed with careful attention to their potential social consequences. A wave of fair decision-making algorithms and more particularly fair machine learning models for prediction has been proposed in recent years years\cite{Fair-accurate-education,AIunfairness-education}. Nevertheless, most of the proposed research either deals with inequality in the pre-processing or post-processing steps or considers a model-based in-processing approach. To take any of the aforementioned routes for bias mitigation, it is critical to audit the unfairness of the outcome of the predictive algorithms and identify the most severe unfairness issues to address. 

Following these concerns, fairness audits of algorithmic decision
systems have been pioneered in a variety of fields \cite{kondmann2021under,kearns2018preventing}. The auditing process of unfairness detection of the model provides a comprehensive guideline for education researchers and officials to evaluate the inequalities of predictive modeling algorithms from different perspectives before deploying them in practice. 

In this paper, we first study if predictive modeling techniques for student success show inequalities for or against a sample of marginalized
communities. We use a real national-level education dataset to analyze the case
of discrimination. We consider a wide range of Machine Learning models for student-success predictions. Then, we audit if
prediction outcomes are discriminating against certain subgroups considering different notions of fairness to identify a potential bias
in predictions. Furthermore, we investigate the impact of imputing the missing values using various techniques on model performance and fairness to key insights for educational practitioners for responsible ML pipelines. 
This study has the potential to significantly impact the practice of data-driven decision-making in higher education investigating the impact of a critical pre-processing step on predictive inequalities.
In particular, how imputation impacts the performance and the fairness of a student-success prediction outcome.

\new{In this paper, we present a prospective validation of predictive modeling on operational data as an important step in assessing the real-world performance of machine learning models. We conduct a prospective evaluation for Out-Of-Sample performance evaluation whereby a model learned from historical data is evaluated by observing its performance on new data. Prospective evaluation is likely to provide a less biased estimation of future performance than the evaluation of historical data.
Ensuring that a model continues to perform well for integration into decision-making workflows and trustworthy operational impact, we simulate testing observations by adding subtle noise to a different group of predictors. To this end, we perturb the sensitive attribute \textit{race}, non-sensitive attributes, and all predictors in three different scenarios to demonstrate external generalization capacity to the community.  
}

We predict the most common proxy attribute \emph{graduation completion} concerning equal treatment to different demographic groups through different notions of fairness. The comprehensive study of the real large-scale dataset of ESL:2002 allows us to validate the performance of different ML techniques for predictive analytics in higher education in a real situation. 
To the best of our knowledge, none of the existing fair machine learning (ML) models have studied existing large-scale datasets for student success modeling. Most of the extant applications of fair ML demonstrate results using small datasets considering a limited number of attributes (e.g., \cite{kleinberg2018algorithmic,yu2020towards}) or in a specific context, such as law school admission \cite{kusner2017counterfactual,cole2019avoiding}.

\section{Bias in Education}
\emph{``Bias in, bias out''.}
The first step toward auditing and addressing disparity in student success prediction is to understand and identify different sources of bias in the dataset. Most of the social data including education data are almost always biased since they inherently reflect historical biases and stereotypes \cite{olteanu2019social}. Data collection and representation methods often introduce additional bias. Disregarding the societal impact of modeling the biased data, further exacerbates the discrimination in the predictive modeling outcome. The term bias refers to demographic disparities in the sampled data that compromise its representativeness \cite{olteanu2019social,fairmlbook}. \emph{Population bias} in the data prevents a model to be accurate for minorities \cite{asudeh2019assessing}. \new{Table 1 presents the racial population bias in the ELS \footnote{Education Longitudinal Study (ELS:2002) is a nationally representative, longitudinal study of 10th graders in 2002 and 12th graders in 2004. \url{https://nces.ed.gov/surveys/els2002/}} dataset for students who attended four-year institutions. Therefore, students who identify as ``White'' are considered as the majority accounting for 66\% of the observations, while ``Multiracial''\footnote{We refer to students with two or more races as multiracial and denote it as MR}, ``Hispanic'' and ``Black'' groups are underrepresented, making them minorities.}

\begin{table}[htbp]
\centering
\scriptsize
\begin{adjustbox}{width=0.5\textwidth}
    \begin{tabular}{|l|r|}
    \toprule
    \multicolumn{1}{|c|}{\textbf{Race}} & \multicolumn{1}{c|}{\textbf{Percent of Population}} \\
    \midrule
    Asian & 11.13\% \\
    \midrule
    Black & 10.21\% \\
    \midrule
    Hispanic & 8.13\% \\
    \midrule
    Multiracial & 4.27\% \\
    \midrule
    White & 65.83\% \\
    \bottomrule
    \end{tabular}%
  \end{adjustbox}
  \label{tab:ELS_pop}%
  \caption{ELS population bias}
\end{table}%

\begin{figure}[!tb]
\centering    
\subfigure[Degree Attainment
]{\label{fig:attain1}\includegraphics[width=0.47\linewidth]{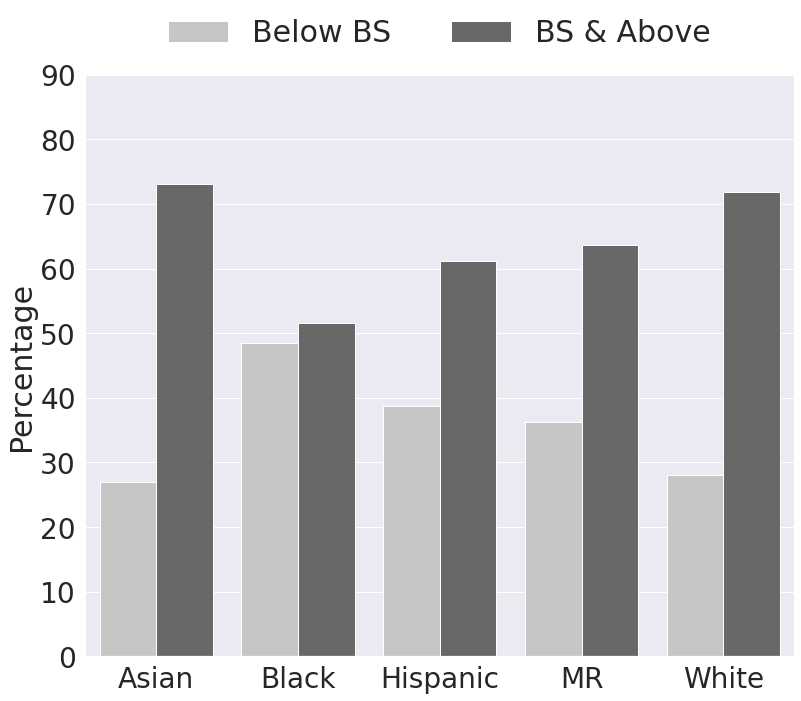}}
\subfigure[Attainment vs. Family Income]{\label{fig:attain2}\includegraphics[width=0.52\linewidth]{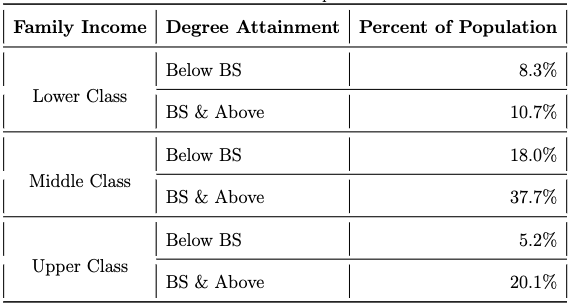}}
\caption{ELS behavioral bias}
\label{fig:attain}
\end{figure}

\new{On the other hand, bias exists in the distribution of attributes' values across different demographic groups, which is referred to as \emph{behavioral bias}. Bias in the data has a direct impact on the algorithmic outcomes. One explanation is the strong correlation between sensitive attributes with other attributes and the response. Figure~\ref{fig:attain1} shows the behavioral bias of the highest degree earned by different racial groups. The histogram indicates that among Black and Hispanic communities, degree attainment below the bachelor's level is very frequent. We can observe another example of behavioral bias in Figure~\ref{fig:attain2}, where students from middle-class and low-income families have degree attainment lower than bachelor's level. Therefore, using degree attainment as a student success indicator calls for careful action and thought to reduce the effects of both population and behavioral bias.}

Transitioning toward a more comprehensive and specified model of disparities among population groups, yields a greater understanding of the key drivers of disparities among population groups.
The key causes behind disparities among population groups that have been identified in the previous education research include, but are not limited to, social class\cite{disparity-racial,disparity-social}, race and ethnicity \cite{disparity-racial}, gender \cite{disparity-structure,disparity-gender2,voyer2014gender}, household characteristics(e.g. education of adults) \cite{disparity-structure}, community characteristics (e.g. presence of schools) \cite{disparity-structure}, and socioeconomic status\cite{disparity-structure,disparity-socioeconomic}.

\new{In this paper, we aim to investigate the impact of the data pre-processing step on disparities in the prediction outcome. Bias detection sheds light on identifying the correct data pre-processing and imputation technique to alleviate the adverse impact of imputation on the performance and fairness of the model. To accomplish this end, we audit the unfairness of the predictive outcome before and after imputation using different scenarios that can occur in practice. We demonstrate how the obtained unfairness results differ and discuss the logic and implications behind each strategy.}
By examining bias in existing data and identifying the key characteristics of vulnerable populations, this paper illuminates how predictive models can produce discriminatory results if the bias is not addressed, and how we need to control the predictive outcome disparity.

\begin{figure}[!tb]
\centering     
\subfigure[]{\label{fig:box1}\includegraphics[width=0.45\linewidth]{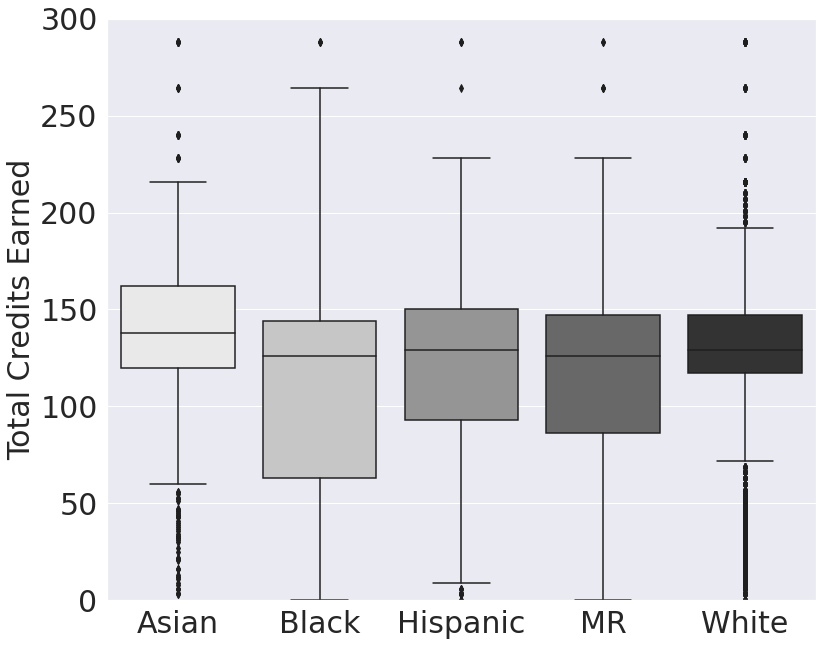}}
\subfigure[]{\label{fig:box2}\includegraphics[width=0.45\linewidth]{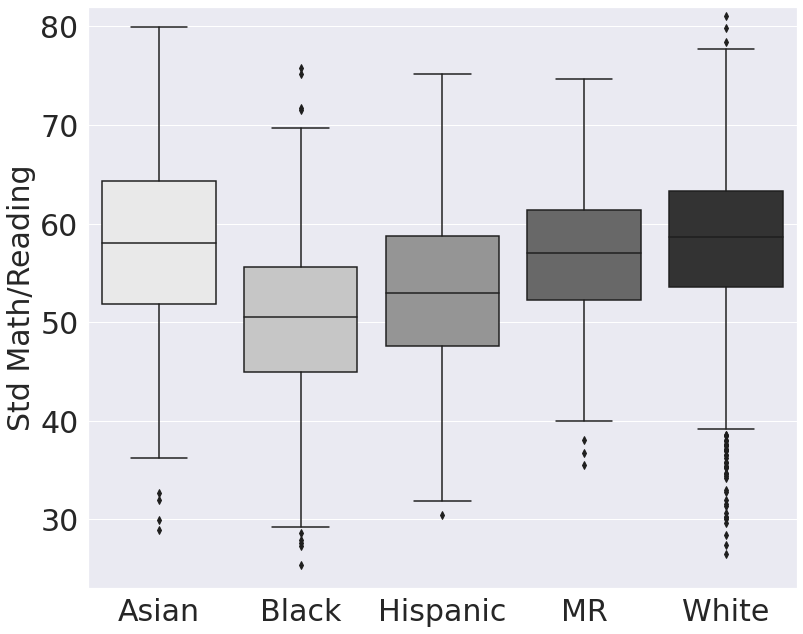}}
\subfigure[]{\label{fig:box3}\includegraphics[width=0.45\linewidth]{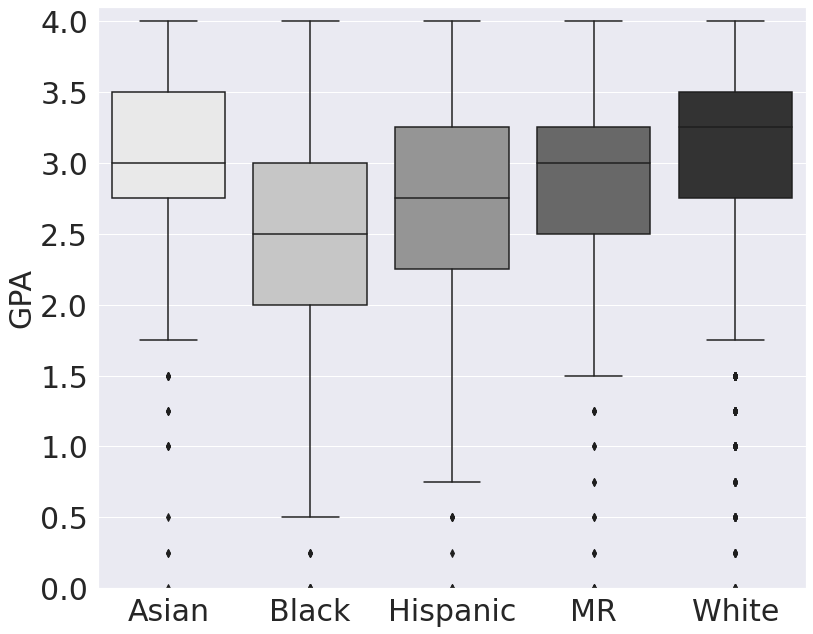}}
\caption{Boxplots of unprotected attributes $\not\!\perp\!\!\!\perp$ race}
\label{fig:boxplots}
\end{figure}

To identify potential sources of behavioral bias, Figure \ref{fig:box1}, \ref{fig:box2} and \ref{fig:box3} illustrate racial disparities with respect to total credits, math/reading test scores, and GPA respectively. More specifically, Figure \ref{fig:box1} shows that Black and Hispanic groups have lower median earned credits with their first and second quartile (50\% of observations) plotted with lower values compared to others. Similarly, Figure \ref{fig:box2} indicates that the student standardized combined math/reading test score has a lower median for Black and Hispanic groups. Moreover, the GPA score of vulnerable students (Black and Hispanic) have a lower median as shown in Figure~\ref{fig:box3}. In addition, the size of each boxplot (from lower quartile to upper quartile) provides insight into the distribution of each group. For example, in Figure \ref{fig:box1}, the Black subgroup has a large box plot meaning that these students have very different outcomes in terms of total earned credits from very low to high values. However, the box plot for the White group of students indicates more similar credit outcomes, mainly distributed around the median value. 

\section{Fairness In Predictive Modeling}\label{sec:fair-predictive}

\emph{Fairness-aware learning} has received considerable attention in the machine learning literature (fairness in ML) \cite{fairML3,fairML4-zafar}. More specifically, fairness in ML seeks to develop methodologies such that the predicted outcome becomes fair or non-discriminatory for individuals based on their protected attributes such as race and sex. 
The goal of improving fairness in learning problems can be achieved by intervention at pre-processing, in-processing (algorithms), or post-processing strategies. 
Pre-processing strategies involve the fairness measure in the data preparation step to mitigate the potential bias in the input data and produce fair outcomes~\cite{feldman2015certifying,kamiran2012data,calmon2017optimized}.
In-process approaches~\cite{zafar2015fairness, zhang2021omnifair, anahideh2020fair} incorporate fairness in the design of the algorithm to generate a fair outcome. Post-process methods~\cite{pleiss2017fairness,feldman2015certifying,zehlike2017fa}, manipulate the outcome of the algorithm to mitigate the unfairness of the outcome for the decision-making process.

There are various definitions for fairness in the literature~\cite{vzliobaite2017measuring,fairmlbook,narayanan2018translation,barocas2016big,dwork2012fairness}. 
The fairness definitions fall into different categories including 
Statistical Parity~\cite{hardt2016equality},
Equalized Odds~\cite{hardt2016equality}, Predictive Equality~\cite{corbett2017algorithmic} and Equal Opportunity~\cite{madras2019fairness}. Table \ref{tab:notions} demonstrates the mathematical definitions of each of these common metrics \cite{makhlouf2021applicability}. Let $S=\{0,1\}$ be a binary sensitive attribute, in a binary classification setting let $Y=\{0,1\}$ be the true label, and $\hat{Y}=\{0,1\}$ be the predicted class label. Most of the fairness notions are derived based on conditional probabilities using these variables to reveal the inequalities of the predictive model. 
Evaluating the fairness of algorithmic predictions requires a notion
of fairness, which can be difficult to choose in practice. Different metrics have been leveraged regarding different contexts, business necessities, and regulations. 
A predictive modeling outcome might have inequalities under one notion of fairness and might not have any under others.

\begin{table}[!tb]
\begin{adjustbox}{width=1\columnwidth}
  \centering
    \begin{tabular}{|c|c|}
    \toprule
    \textbf{Fairness Notion } & \textbf{Formulation } \\
    \midrule
    Statistical Parity ({\bf SP})  & $P(\hat{Y}=1|S=1)-P(\hat{Y}=1|S=0)$ \\
    \midrule
    Equalized Odds ({\bf EO}) & $P(\hat{Y}=1|Y=y,S=1)-P(\hat{Y}=1|Y=y,S=0),  \forall {y \in \{0,1\}}$ \\
    \midrule
    Equal Opportunity ({\bf EoP})&  $P(\hat{Y}=1|Y=1,S=1)-P(\hat{Y}=1|Y=1,S=0)$
    \\
    \midrule
    Predictive Equality ({\bf PE})&  $P(\hat{Y}=1|Y=0,S=1)-P(\hat{Y}=1|Y=0,S=0)$
    \\
    \midrule

    \end{tabular}%
    \end{adjustbox}
    \caption{Common Fairness Definitions}
  \label{tab:notions}%
\end{table}%

In the education context, to give some examples, a) Demographic (statistical) parity is referred to as the \emph{discrepancy of the predicted highest level of degree (success) across different demographic groups of students}, and  b) Equal opportunity indicates the \emph{discrepancy of the predicted highest level of a degree across different demographic groups of students, given their success, is 1}. In this paper, we use a binary classification setting but across multilevel racial and gender population subgroups ($S$ is not necessarily binary). 

We extend the fairness metrics described in Table~\ref{tab:notions} for non-binary sensitive attributes, by considering \emph{one-versus-rest} approach for unfairness calculation. More specifically, to calculate the unfairness gaps, we consider each subgroup as $S=1$ and compare it against the rest $S=0$ (i.e. all other subgroups), one at a time. In this paper, we mainly focus on racial disparities, however, our proposed approach for auditing fairness and investigating the imputation impact can be extended to use other sensitive attributes. For example, the decision maker can use "gender" as a sensitive attribute.

\section{Fairness Audits}

Notwithstanding the awareness of biases and unfairness in machine learning, the actual challenges of ML practitioners have been discussed in a few previous research \cite{veale2018fairness,holstein2019improving} with the focus on specific contexts, such as predictive policing \cite{propublica} and
child mistreatment detection \cite{chouldechova2018case}. 
ML practitioners often struggle to apply existing auditing and de-biasing methods in their contexts \cite{holstein2019improving}. The concept of auditing algorithms and ethics-based auditing in various contexts has lately reached its pinnacle. \cite{mokander2021ethics,raji2020closing,wilson2021building,kondmann2021under,mokander2021ethics}. The final goal of the fairness auditing process is to determine whether the ML model's results are fair. As a result, the auditing process aids in determining the appropriate actions to take, the best bias mitigation method to employ, and the most suitable technique to use throughout the ML development pipeline \cite{raji2020closing}.

A designated user of predictive modeling in higher education needs support to audit the ML model performance and inequalities before adopting and deploying it in practice. 
To address the education practitioners and policymakers on assessing the inequalities of the predictive outcome, in this paper, we audit the unfairness of ML models for student success prediction using major notions of fairness to identify bias in algorithms using different metrics, Table \ref{tab:notions}. We also audit unfairness to ensure an ethical pre-processing approach. We audit a wide range of fairness metrics and conduct a comprehensive analysis of the performance of different ML models and their inequalities across different racial and gender subgroups throughout the data preparation (imputation) and the model training steps using the ELS dataset.


\vspace{-2mm}
\section{Success Prediction} 
Before moving to the ML pipeline, we first discuss the prediction problem of interest. In this paper, we specifically focus on predicting the academic success of students in higher education. Student-success prediction is critical for institution performance evaluation, college admission, intervention policy design, and many more use cases in higher education \cite{yu2020towards,stephan2015will}.

Quantifying student success is a very complex topic since the true quality of any candidate is hidden and there is limited information available. There are proxy attributes such as first-year GPA or graduation completion that are typically used as the measure of success. In this work, we are primarily interested in studying the prediction of the highest level of degree (classification problem) using the ELS:2002 dataset. 

Numerous factors can affect student success \cite{voyer2014gender}. Thus, identifying the most informative and significant subset of potential variables is a critical task in predictive modeling \cite{FSstudy}. To select a proper subset of attributes, we conducted a thorough literature search and combine it with the domain expert knowledge. These factors include, but are not limited to, academic performance (SAT scores, GPA) \cite{chamorro2008personality}, student demographic attributes (e.g. race, gender) \cite{voyer2014gender,fairethic-education}, socio-economic status \cite{disparity-structure,disparity-socioeconomic2}, environmental factors, and extra(out of school) activities.  

Incorporating protected attributes in the modeling procedure has raised concerns in the fair-ML domain \cite{barocas2016big}. 
The predictive outcome depends on the information available to the model and the specific algorithm used. A model may leverage any feature associated with the outcome, and common measures of model performance and fairness will be essentially unaffected. In contrast, in some cases, the inclusion of unprotected attributes may adversely affect the performance and fairness of a model due to a latent correlation with other protected attributes. In this paper, we audit the unfairness of the model and the impact of imputation incorporating the sensitive attribute as the determinant.
Decision Tree \cite{DT-corr-perform}, Random Forest \cite{RF-risk}, K-Nearest Neighbor \cite{dudani1976distance} \cite{tanner2010predicting}, Logistic Regression \cite{thompson2018predicting}, and SVM \cite{SVM-DT-NN} are among the well-known ML models in higher education. Table \ref{tab:Missings}, represents the list of variables in this study, and their corresponding missing value percentages.

\begin{table}[!tbh]
\scriptsize
  \centering
    \begin{adjustbox}{width=1\columnwidth}
   \begin{tabular}{|l|c|l|c|}
    \toprule
    \textbf{Variables } & \textbf{\% Missing} & \textbf{Variables } & \textbf{\% Missing} \\
    \midrule
    S-T relationship & 33.32 & Std Math/Reading & 0.02 \\
    \midrule
    F3-loan-owed & 25.33 & F3\_Employment & 0 \\
    \midrule
    \%white teacher & 23.69 & F3\_Highest level of education & 0 \\
    \midrule
    \%Black teacher & 19.85 & High school attendance & 0 \\
    \midrule
    \%Hispanic teacher & 17.72 & Family Composition & 0 \\
    \midrule
    TV/video(h/day) & 14.87 & Race\_Hispanic, race specified & 0 \\
    \midrule
    Work(h/week) & 12.06 & F3\_Separated no partner & 0 \\
    \midrule 
    F2\_College entrance & 9.75 & F3\_Never Married w partner & 0 \\
    \midrule
    Generation & 7.06 & F3\_Never Married no partner & 0 \\
    \midrule
    F3\_GPA(first attended) & 6.79 & F3\_Married & 0 \\
    \midrule
    F3\_GPA(first year) & 6.78  & F3\_Divorced/Widowed w partner & 0 \\
    \midrule
    F1\_TV/video(h/day) & 6.76 & F3\_Divorced/Widowed no partner & 0 \\
    \midrule
    F1\_units in math & 6.13 & Race\_White & 0 \\
    \midrule
    Athletic level & 5.39  & Race\_More than one race & 0 \\
    \midrule
    F1\_frequency of computer use & 4.27  & Race\_Hispanic, no race specified & 0 \\
    \midrule
    Credits (total)  & 4.04  & School Urbanicity & 0 \\
    \midrule
    F1\_Std Math & 3.64  & Race\_Black or African Amer & 0 \\
    \midrule
    Credits (first year) & 3.48  & Race\_Asian, Hawaii/Pac. Isl & 0 \\
    \midrule
    F3\_GPA (all)	 & 3.33  & Race\_Amer. Indian/Alaska & 0 \\
    \midrule
    F3\_Credits\_Math & 3.01  & Gender\_Male & 0 \\
    \midrule
    F3\_Credits\_Science & 2.90 & Gender\_Female & 0 \\
    \midrule
    F1\_Work(h/week) & 2.77  & Parents education & 0 \\
    \midrule
    Homework(h/week) & 1.85  & Income & 0 \\
    \midrule
    Number of school activities & 0.84   & F1\_Drop out & 0 \\
    \midrule
    English	 & 0.02  & F3\_Separated w partner & 0 \\
    \bottomrule
    \end{tabular}%
    \end{adjustbox}
  \caption{List of Variables} 
    \label{tab:Missings}%
\end{table}%

 \clearpage

\subsection{Missing Values and Imputation}\label{sec:data-preprocessign}

\emph{Missing Values} are the frequent latent causes behind many data analysis challenges, from modeling to prediction, and from accuracy to fairness for protected (sensitive) subgroups. Therefore, \emph{handling Missing values} is a complicated problem that requires careful consideration in education research \cite{missing-review1,Missing3-MI}. In this regard, different imputation techniques have been proposed in the literature and the effectiveness of each methodology on various applications has been studied. \emph{Mean Imputation}, \emph{Multiple Imputation}, and other clustering-based imputation strategies such as \emph{KNN-imputation}, are among the well-known techniques in handling missing values which we will briefly describe here. 

Most large-scale and nationally representative education data sets, (e.g., ELS) suffer from a significant number of incomplete responses from the research participants. While features that contain more than 75\% missing or unknown values are not usually informative, most features suffer from less than 25\% missing values and are worth keeping. Removing all observations with missing values induces significant information loss in success prediction.

\textbf{Simple Imputation} is one the most basic imputation strategies. The process involves replacing the missing variable of observation with the mean (or median) of the observations with available values for the same variable. The mean imputation method is known to decrease the standard error of the mean. This fact exacerbates the risk of failing to capture the reality through statistical tests
\cite{missing-book,missing-review1}.
\textbf{Multiple Imputation (MI) \cite{rubin1996multiple}} is a more advanced imputation strategy that aims to estimate the natural variation in the data by performing several missing data imputations. In fact, MI produces a set of estimates through various imputed datasets and combines them into a single set of estimates by averaging across different values. The standard errors of parameter estimates produced with this
the method has shown to be unbiased \cite{rubin1996multiple}.
\textbf{KNN Imputation} is a non-parametric imputation strategy, which has been shown to be successful for different contexts \cite{missing-knn}. KNN imputer replaces each sample’s missing values with the mean value from $K$ nearest neighbors found in the dataset. In fact, two samples are considered close neighbors if the features that neither is missing are close. KNN imputation is able to capture structure in the dataset while the underlying data distribution is unknown \cite{somasundaram2011evaluation}. To the best of our knowledge, KNN imputation has not been used in the education context.

Overall, ignoring missing data can not be an effective approach of handle missing values, and more importantly, can result in predictive disparity for minorities. While many education-related studies addressed the challenges of missing data, as discussed, little is known about the impact of applying different imputation techniques on the fairness outcome of the final model. In this project, we aim to address this gap by considering the three above-mentioned imputation strategies. To the best of our knowledge, none of the prior works in the education domain worked on ensuring fairness while imputing missing values in pre-processing steps.

\vspace{-3mm}
\section{Experiments}

\stitle{Data Preparation.} As previously stated, we use the ELS dataset in this study to audit the fairness of ML models in the development pipeline for predicting student success.
The ELS dataset includes many categorical variables. Therefore, we begin by creating appropriate labeling and converting categorical attributes to numeric ones (dummy variable) following the NCES dataset documentation\footnote{
\url{https://nces.ed.gov/surveys/els2002/avail_data.asp}}).
Next, we perform a transformation on the considered response variable \emph{highest level of degree} to construct a binary classification problem. That is, we label students with a college degree (BS degree and higher) as the favorable outcome (label=1), and others as the Unfavorable outcome (label=0). Note that the dataset is filtered based on the institution type to only include students who attended four-year postsecondary institutions. We consider ``race'' as the sensitive attribute for fairness evaluation, which includes five groups of White, Black, Hispanic, Asian, and Multi-racial(MR) students.  
A Data cleaning is performed to identify and rename the missing values (based on the documentation) and remove the observations that have many missing attributes ($>75\%$ of the attributes are missing).

\stitle{Imputation.} The final and significantly important task is then to handle the remaining missing values in the dataset. 
We consider different imputation techniques; Simple Imputation {\bf SI}, Multiple Imputation {\bf MI}, and KNN Imputation {\bf KNN-I}. We consider a baseline where we remove observations with missing attributes and refer to it as {\bf Remove-NA}.

\stitle{Model Training} procedure follows the data preparation step as we obtain clean-format datasets. We aim to analyze the performance of different Machine learning (ML) models under each imputation technique and data split scenario, to audit the inequalities in the prediction outcome. We consider Decision Tree {\bf DT}, Random Forest {\bf RF}, Support Vector Classifier {\bf SVC}, and Logistic Regression {\bf Log} ML models.

\subsection{Prospective Evaluation}

\new{To investigate the impact of imputation and data representation on the accuracy and fairness of the outcome of the predictive modeling, we simulate various scenarios to provide a trustworthy evaluation of models and their generalization capacity for real-time deployment. Table~\ref{tab:Scenarios} summarizes our considered scenarios to evaluate imputation impact on the model unfairness.
\textbf{RNA.rnd} indicates the common practice data pre-processing step in the machine learning literature where all rows with missing values are removed before a train/test split. Similarly, in \textbf{RNA.str} we remove the missing values while performing stratification on the ``race'' and ``response'' (highest level of degree) variables to ensure that students from each racial subgroup with different success outcomes (response) are well-represented in both training and testing datasets.} 

\begin{table}[h!]
\scriptsize
  \centering
   \begin{adjustbox}{width=1\columnwidth}
    \begin{tabular}{| l | l |  p{7cm} |}
    \hline
    Scenario & Missing Values & Train Test Data \\ \hline
    RNA.rnd & Removed & 80:20 split  \\ \hline
    RNA.str & Removed & 80:20 split, stratified on race and target variable  \\ \hline
    Imp.rnd & Imputed & 80:20 split  \\
    \hline
    Imp.str & Imputed & 80:20 split, stratified on race and target variable  \\
    \hline
    Imp.prop & Imputed & 80:20 split, dataset separated based on racial groups, where training/testing data is proportional to the fraction of observations within each group  \\
    \hline
    Imp.prop.frac & Imputed & Similar to Imp.prop scenario but only 75\% of the testing data is chosen (randomly) for testing  \\
    \hline
    Imp.prop.frac.perturb & Imputed & Similar to Imp.prop.frac where the dataset is also simulated by performing three different perturbations on race, nonsensitive, and all attributes \\
    \hline
    \end{tabular}%
     \end{adjustbox}
\caption{Evaluation Scenarios}
\label{tab:Scenarios}%
\end{table}%

\new{In \textbf{Imp.rnd}, we split the entire dataset into train/test sets, perform an imputation technique on the training set, and transfer the trained imputer on the testing set to replace missing values. In \textbf{Imp.str} we perform imputation similar to \textbf{Imp.rnd}, however, we consider stratification on the ``race'' and ``response'' variables to generate representative train and test datasets. In another attempt to maintain the distribution of different population subgroups, the \textbf{Imp.prop} scenario by fixing the fraction of observation from each racial group in train/test splits to the fractions in the entire dataset. 
To simulate a real-time scenario, we randomly select a smaller subset of the remaining observations for testing in \textbf{Imp.prop.frac} since there is no guarantee that future observations to maintain the same distribution as training data. Lastly, in \textbf{Imp.prop.frac.perturb}, a modified testing dataset is generated through perturbation of different groups of attributes to simulate the representation of different racial groups and their associated attribute values for future observations. That is, we perturb ``race'', ``non-sensitive'' attributes, and ``all attributes'', respectively.
}

\subsection{Results}

\new{In this section, we summarize our findings in three key discussions. First, we compare the performance and unfairness of the considered ML models using different imputation strategies. Next, we compare random and stratified imputed data against {\bf Remove-NA} scenarios. We then simulate different perturbation scenarios to study the impact of imputation on the unfairness while considering unknown distribution for future observations (real-time scenario). Lastly, we compare different pre-processing approaches and real-time scenarios to offer general  guidelines for education practitioners.}

\begin{figure*}[!t]
\centering  
\subfigure{\label{exp:alpha-compas}\includegraphics[width=0.7\linewidth]{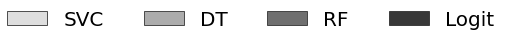}}
\subfigure[Remove-NA]{\label{exp:alpha-compas}\includegraphics[width=0.45\linewidth]{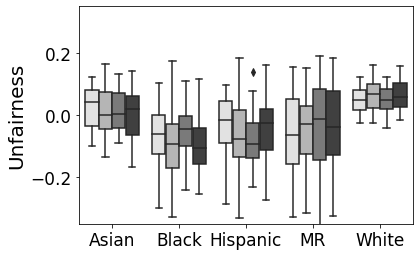}}
\subfigure[Simple Imp.]{\label{exp:nested-compas}\includegraphics[width=0.45\linewidth]{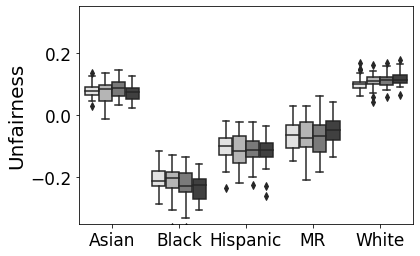}}
\subfigure[Multiple Imp.]{\label{exp:append-compas}\includegraphics[width=0.45\linewidth]{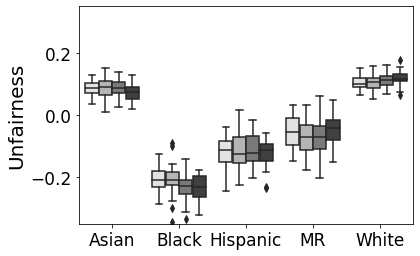}}
\subfigure[KNN Imp.]{\label{exp:alpha-adult}\includegraphics[width=0.45\linewidth]{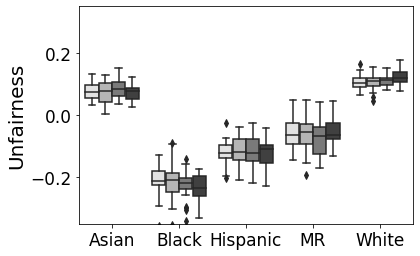}}
\caption{Impact of Imputation - Statistical Parity }\label{exp:imp_sp}
\vspace{-4mm}
\end{figure*}

\new{Figure~\ref{exp:imp_sp} reveals the impact of utilizing different imputation techniques on the unfairness of different ML models based on the Statistical Parity (\textbf{SP}) notion. We can observe that the unfairness has increased for Black and Hispanic groups of students after imputation. It is worth mentioning that imputation significantly reduces the variance of unfairness across different racial groups as it involves more observations.
The unfavorable unfairness hike is still true for all ML models, if we solely change the imputation strategy. As a result, even while imputation has a considerable impact on the level of unfairness (raising the unfairness mean and decreasing the variance), the imputation techniques evaluated in this study are barely different from one another. Additionally, the impact of various imputation methods on unfairness exhibits a consistent pattern based on other fairness notions presented in \S~\ref{sec:appendix}.}

\begin{figure*}[!t]
\centering  
\subfigure{\label{exp:alpha-compas}\includegraphics[width=0.7\linewidth]{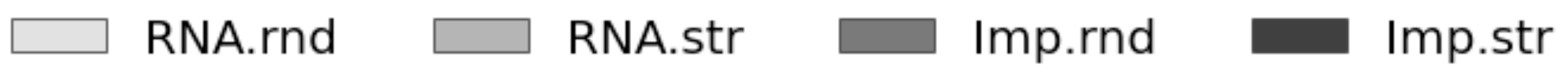}}
\subfigure[Statistical Parity]{\label{exp:alpha-compas}\includegraphics[width=0.49\linewidth]{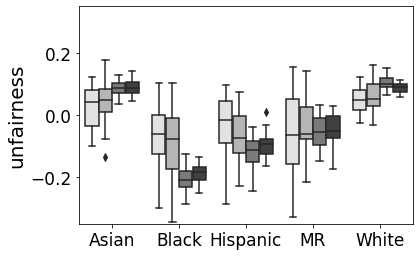}}
\subfigure[Predictive Equality]{\label{exp:nested-compas}\includegraphics[width=0.49\linewidth]{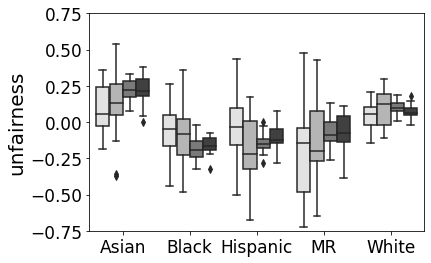}}
\subfigure[Equal Opportunity]{\label{exp:append-compas}\includegraphics[width=0.49\linewidth]{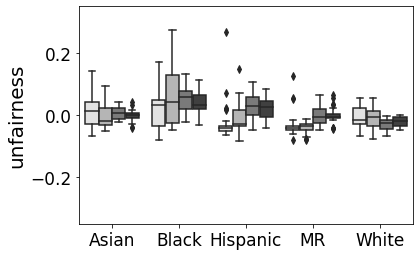}}
\subfigure[Equalized Odds]{\label{exp:alpha-adult}\includegraphics[width=0.49\linewidth]{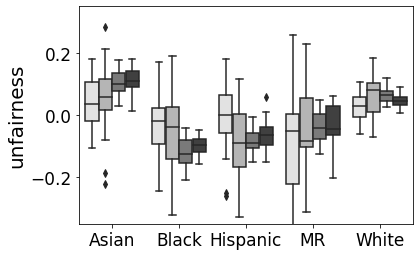}}
\subfigure[Testing Accuracy]{\label{exp:alpha-adult}\includegraphics[width=0.49\linewidth]{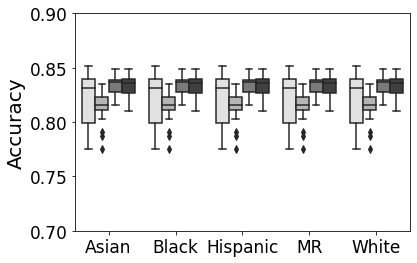}}
\caption{Remove-NA vs. Imputation (SVC)}\label{exp:RNA_Imp}
\vspace{-4mm}
\end{figure*}

\new{Figure~\ref{exp:RNA_Imp} presents a comparison between  \textbf{RNA-str} and \textbf{RNA-rnd} with imputing the missing values with \textbf{Imp-str} and without \textbf{Imp-rnd} stratification. Although imputation improves the prediction accuracy, it hurts the unfairness gap for Black and Hispanic groups. For example, the average unfairness based on \textbf{SP} increases in magnitude for both Black and Hispanic students after imputation (\textbf{Imp-str} and \textbf{Imp-rnd} ) compared to Remove-NA scenarios (\textbf{RNA-str} and \textbf{RNA-rnd}). One justification behind this observation can be related to the size of the testing set, which significantly increases after imputation, Table~\ref{tab:popsize}. 
The average unfairness gaps increase as we include more observations from the minority (underprivileged) group of students via imputation.}

\begin{table}[h!]
\scriptsize
  \centering
    \begin{tabular}{|r|l|r|r|r|r|}
    \toprule
    \rowcolor[rgb]{ .929,  .929,  .929} \multicolumn{1}{|c|}{\textbf{Scenario}} & \multicolumn{1}{c|}{\textbf{Race}} & \multicolumn{1}{c|}{\textbf{ Pop. $X_{train}$}} & \multicolumn{1}{c|}{\textbf{ Pop. $X_{test}$}} & \multicolumn{1}{c|}{\textbf{ $Y=1$}} & \multicolumn{1}{c|}{\textbf{ $\hat{Y}=1$}} \\
    \midrule
          & Asian & 95  & 26  & 20   & 22 \\
\cmidrule{2-6}          & Black & 75   & 19   & 13   & 15 \\
\cmidrule{2-6}    \multicolumn{1}{|l|}{\textbf{RNA.rnd}} & Hispanic & 75   & 17   & 13   & 14 \\
\cmidrule{2-6}          & Multiracial & 32   & 9   & 7   & 8 \\
\cmidrule{2-6}          & White & 712  & 175  & 128  & 145 \\
    \midrule
          & Asian & 494  & 123  & 90   & 108 \\
\cmidrule{2-6}          & Black & 452  & 114  & 62   & 77 \\
\cmidrule{2-6}    \multicolumn{1}{|l|}{\textbf{Imp.rnd}} & Hispanic & 357   & 94   & 58   & 67 \\
\cmidrule{2-6}          & Multiracial & 190   & 47   & 30   & 35 \\
\cmidrule{2-6}          & White & 2923  & 727  & 526  & 598 \\
    \midrule
          & Asian & 494  & 123  & 90   & 100 \\
\cmidrule{2-6}          & Black & 452  & 114  & 58   & 66 \\
\cmidrule{2-6}    \multicolumn{1}{|l|}{\textbf{Imp.prop}} & Hispanic & 361   & 90   & 55   & 65 \\
\cmidrule{2-6}          & Multiracial & 190   & 47   & 30   & 36 \\
\cmidrule{2-6}          & White & 2920  & 730  & 525  & 614 \\
    \bottomrule
    \end{tabular}%
  \label{tab:popsize}%
   \caption{Population size of demographic groups}
\end{table}%

\begin{figure*}[!t]
\centering  
\subfigure{\label{exp:alpha-compas}\includegraphics[width=0.8\linewidth]{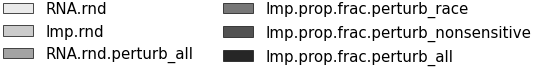}}
\subfigure[Statistical Parity]{\label{exp:alpha-compas}\includegraphics[width=0.49\linewidth]{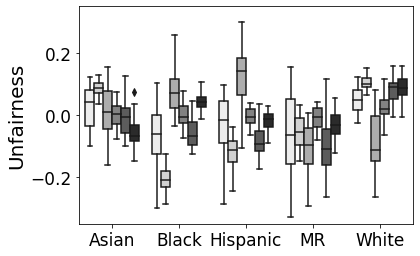}}
\subfigure[Predictive Equality]{\label{exp:nested-compas}\includegraphics[width=0.49\linewidth]{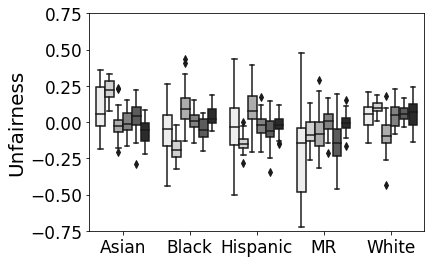}}
\subfigure[Equal Opportunity]{\label{exp:append-compas}\includegraphics[width=0.49\linewidth]{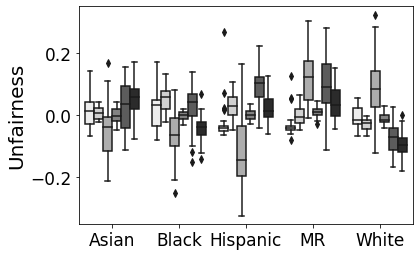}}
\subfigure[Equalized Odds]{\label{exp:alpha-adult}\includegraphics[width=0.49\linewidth]{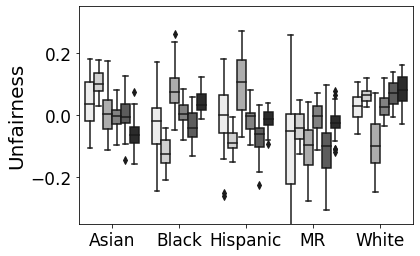}}
\subfigure[Testing Accuracy]{\label{exp:alpha-adult}\includegraphics[width=0.49\linewidth]{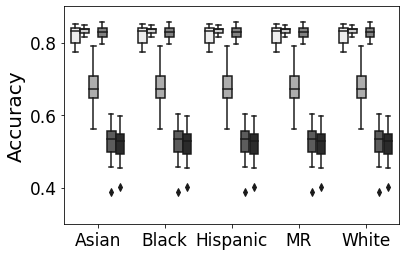}}
\caption{Fairness and Performance Evaluation of Different Perturbation Scenarios compared against Remove-NA and Imp. (SVC)}
\label{exp:all_perturb}
\vspace{-4mm}
\end{figure*}

\new{
Figure~\ref{exp:all_perturb} presents the results of considered unfairness metrics for different evaluation scenarios. Recall that the perturbation scenarios simulate real-time testing data, which may or may not follow the original distribution in the original dataset. That is, we merely perturb the ``race'' in \textbf{Imp.prop.frac.perturb-race}, all non-sensitive attributes (all attributes except race and gender) in \textbf{Imp.prop.frac.perturb-nonsensitive}, and all attributes in \textbf{Imp.prop.frac.perturb-all}. First, we notice that perturbation based on race decreases unfairness, compared to \textbf{RNA-rnd} and \textbf{Imp-rnd}, while increasing the accuracy. This indicates that perturbing the sensitive attribute in the testing data first equalizes the representation of all groups. As a result, it decorrelates the sensitive and non-sensitive attributes inducing unfairness reduction and accuracy improvement. In addition, perturbing the non-sensitive attributes decreases unfairness for Black and Hispanic racial groups while decreasing the model accuracy. Note that unfairness reduction is less significant than perturbing race only. It is possible that only the correlation between sensitive and non-sensitive attributes has been decreased, leaving the representation bias (population bias) unchanged when non-sensitive attributes are perturbed. A similar pattern can be seen in perturbing all attributes (the most realistic case), which will result in a less-accurate and more fair model compared with the \textbf{RNA-rnd} and \textbf{Imp-rnd} scenarios. Lastly, perturbing all of the attributes of the not-imputed dataset, \textbf{RNA.rnd.perturbed-all}, would reduce the unfairness for vulnerable groups, however, it adversely affects privileged groups. That is because the test set for \textbf{RNA.rnd.perturbed-all} scenario suffers from representation bias.
}

\new{The perturbation results shown in Figure~\ref{exp:all_perturb} reflect scenarios where social injustice has been mitigated through effective interventions such as increasing access to more educational resources (e.g. computers and scholarships) for vulnerable population groups to increase their representation in the applicant pool and boost their chances of success (attain bachelor degree). In such an ideal scenario, the model will will treat every group equally. Note that perturb-all and non-sensitive scenarios have low accuracy since distribution of the input attribute space differs from the distribution used to build the model.} 

\begin{figure*}[!t]
\centering  
\subfigure{\label{exp:alpha-compas}\includegraphics[width=0.6\linewidth]{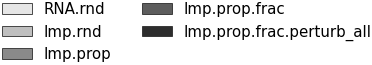}}
\subfigure[Statistical Parity]{\label{exp:alpha-compas}\includegraphics[width=0.49\linewidth]{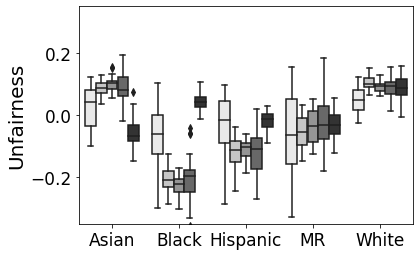}}
\subfigure[Predictive Equality]{\label{exp:nested-compas}\includegraphics[width=0.49\linewidth]{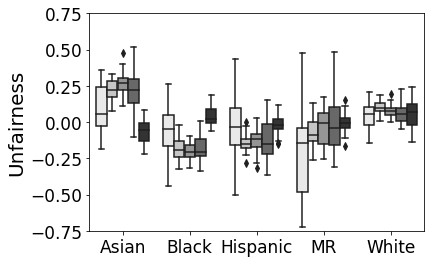}}
\subfigure[Equal Opportunity]{\label{exp:append-compas}\includegraphics[width=0.49\linewidth]{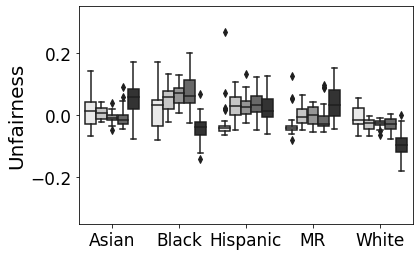}}
\subfigure[Equalized Odds]{\label{exp:alpha-adult}\includegraphics[width=0.49\linewidth]{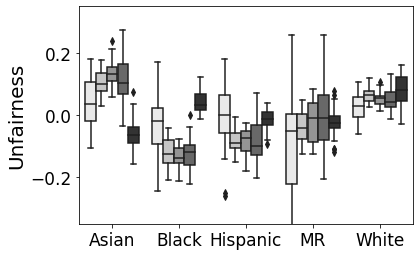}}
\subfigure[Testing Accuracy]{\label{exp:alpha-adult}\includegraphics[width=0.49\linewidth]{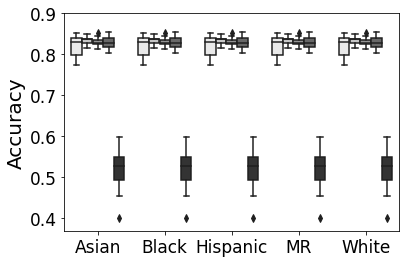}}
\caption{Fairness and Performance Evaluation of Remove-NA, Imputation Scenarios vs. the Perturbation of all attributes (SVC)}\label{exp:all_scenario}
\vspace{-4mm}
\end{figure*}

\new{In Figure~\ref{exp:all_scenario}, we compare the estimated unfairness gaps for perturbation, regular and modified imputation, and compared them against Remove-NA scenarios. First, we note that the unfairness gaps are larger for the imputation scenarios  (\textbf{Imp-rnd} and \textbf{Imp-prop} and \textbf{Imp-prop-frac}) compared to the {\bf RNA-rnd} scenarios. Furthermore, perturbation scenarios show that the high unfairness gaps in imputation scenarios might have been overestimated since the distribution of future observation may differ dramatically from the past if interventions are implemented in society to address the disparity.} \new{By perturbing the data distribution in the test set, we eliminated representation bias and the correlation between sensitive and unprotected attributes.  First, the testing distribution is different than training, and, second, the test set observations exhibit identical characteristics, meaning there is no difference between population subgroups. Note that the model performs similarly poorly (testing accuracy) for all subgroups in this scenario as the testing distribution is different than training (i.e., ``no free lunch theorem'').}

\new{As shown in Table~\ref{tab:notions}, Statistical Parity ({\bf SP}) is a fairness metric that compares positive (or success) prediction outcome ($\hat{Y}=1$) across different racial groups without considering their true $Y$ label (real outcome). Figure~\ref{exp:RNA_Imp}
shows that the unfairness for Black and Hispanic students increases after imputation based on ({\bf SP}), which suggests that fewer students are being predicted as positive (successful) in comparison to other racial groups. On the other hand, Figures~\ref{exp:all_perturb} and ~\ref{exp:all_scenario} indicate that if we perturb all the attributes in the testing data (i.e., enforcing effective interventions), Black and Hispanic students tend to be predicted as positive (successful), thereby reducing the unfairness gap in prediction. 
}

\new{Predictive Equality ({\bf PE}) focuses mainly on unsuccessful students who are incorrectly predicted as successful given their racial subgroup. This type of unfairness could lead to considerably unfavorable results in higher education, where policymakers fail to recognize those in need. Figure~\ref{exp:RNA_Imp} 
reveals that unfairness for Black and Hispanic groups increases after imputation, meaning that more unsuccessful students are incorrectly predicted as positive (successful). Results in Figures~\ref{exp:all_perturb} and ~\ref{exp:all_scenario} show that imputation without consideration of intervention (changing the distribution with perturbation) leads to an almost similar level of unfairness for Black and Hispanic students. However, the unfairness decreases with perturbation/simulated scenario leading to insignificantly small incorrect classification of unsuccessful students from these groups.
}

\new {
Equal Opportunity ({\bf EoP}) also emphasizes the positive prediction outcome and examines how the model correctly classifies successful students given their racial subgroup. Figure~\ref{exp:RNA_Imp} presents that unfairness for Black and Hispanic students does not change considerably with and without imputation since the model correctly classifies the positive class (successful students) as the model accuracy level remains roughly the same. Moreover, Figures~\ref{exp:all_perturb} and ~\ref{exp:all_scenario} further demonstrate that the degree of unfairness based on {\bf EoP} for the perturbation scenario may or may not decrease for Black. The justification lies behind the impact of perturbation on the accuracy level of the model. Changing the distribution reduced the probability of correct classification of successful students (True positives), meaning that they are less correctly classified as successful. Therefore, simulating the race reduces unfairness based on {\bf EoP} through decorrelating sensitive and non-sensitive attributes and improving the generalization power of the model (accuracy level). However, perturbing all the attributes does not necessarily decrease the unfairness gaps for the unprivileged group (Black and Hispanic students) since changing the distribution of all attributes results in less accurate predictions.
}

\new{Equalized Odds ({\bf EO}) measures the True Positive and False Positive rates across different racial groups. It merely emphasizes the positive prediction outcome $\hat{y}=1$, which is \emph{$>$BS degree attainment} conditioned on the true response value and race. Figure~\ref{exp:RNA_Imp}
shows that unfairness for Black and Hispanic students increases after imputation based on ({\bf SP}) which indicates that more students are predicted as positive (successful) compared to other racial groups, which can lead to adverse outcomes if the prediction is a false positive one. On the other hand, Figures~\ref{exp:all_perturb} and ~\ref{exp:all_scenario} indicate that if we perturb all of the attributes in the testing data (real-time case) the Black and Hispanic groups tend to be less predicted as positive (successful) reducing the unfairness and notifying the decision-makers of the required policy changes needed to be considered in future.}

\section{Conclusion}

\new{Education policymakers face numerous challenges, including data preparation when utilizing predictive modeling to aid in decision-making processes. In addition, deploying machine learning on historical data would introduce bias to future predictions. One primary source of bias originates from the representativeness of vulnerable population subgroups in the historical data. In this paper, we assessed the impact of handling missing values with imputation on the prediction unfairness for the large-scale national ELS dataset.}

\new{First, we compared the standard procedure of removing all rows with missing values with the regular train-test split process for imputed missing values. In comparison to the Remove-NA scenario, imputation increases the average while reducing the variance of the unfairness gap for vulnerable groups of students. On the other hand, imputation increases the prediction accuracy of different ML models, indicating a trade-off with fairness. Moreover, we observed that the choice of imputation strategy has an insignificant impact on the unfairness gaps for different population subgroups.}

\new{To evaluate the unfairness of the ML model outcome, we designed various train-test splitting scenarios and altered the distribution of future observations by simulating the testing data. Our analysis indicated that while imputation tends to increase the average unfairness when using the standard training-testing split approach, it assists with attaining fair models when the testing distribution changes for future observations (as a result of effective disparity reduction interventions in society). In other words, if access to educational resources, parental education, community training, and general interventions to support vulnerable students are improved, the upcoming batch of students would have a different distribution. Thereby, the model trained on the imputed training data would be fair enough.}

\newpage
\newpage
 \bibliographystyle{elsarticle-num} 
 \bibliography{ref.bib}
\newpage

\section{APPENDIX}\label{sec:appendix}
\begin{figure*}[hb!]
\centering     
\subfigure{\label{exp:alpha-compas}\includegraphics[width=0.7\linewidth]{figs/fig3_Legend.png}}
\subfigure[Remove NA]{\label{}\includegraphics[width=0.45\linewidth]{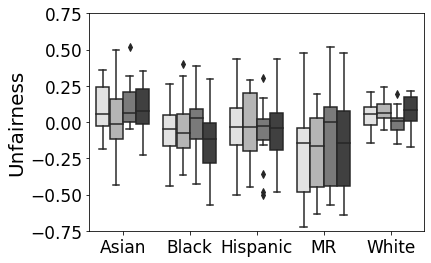}}
\subfigure[Simple Imp.]{\label{}\includegraphics[width=0.45\linewidth]{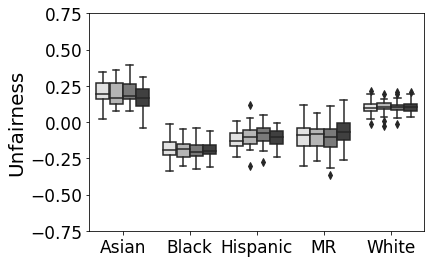}}
\subfigure[Multiple Imp.]{\label{exp:append-compas}\includegraphics[width=0.45\linewidth]{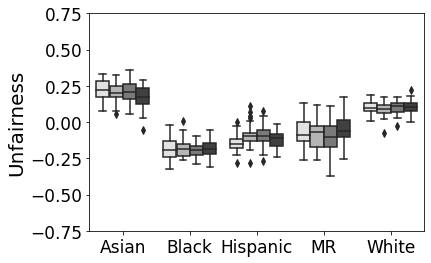}}
\subfigure[KNN Imp.]{\label{}\includegraphics[width=0.45\linewidth]{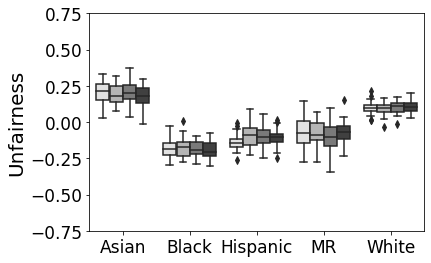}}
\caption{Impact of Imputation - Predictive Equality}\label{exp:bars_pe}
\end{figure*}

\begin{figure*}[hb!]
\centering     
\subfigure{\label{exp:alpha-compas}\includegraphics[width=0.7\linewidth]{figs/fig3_Legend.png}}
\subfigure[Remove NA]{\label{}\includegraphics[width=0.45\linewidth]{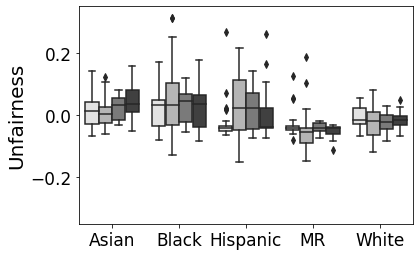}}
\subfigure[Simple Imp.]{\label{}\includegraphics[width=0.45\linewidth]{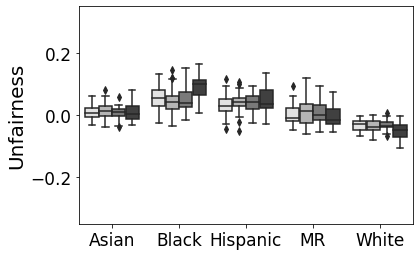}}
\subfigure[Multiple Imp.]{\label{}\includegraphics[width=0.45\linewidth]{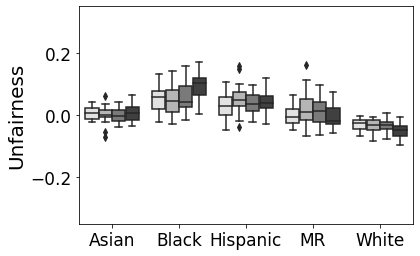}}
\subfigure[KNN Imp.]{\label{}\includegraphics[width=0.45\linewidth]{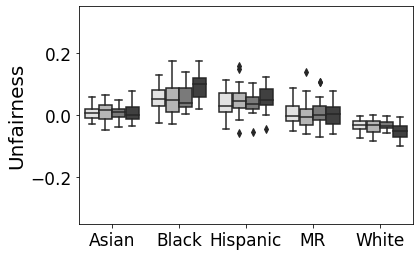}}
\caption{Impact of Imputation- Equal Opportunity}\label{exp:bars_eop}
\end{figure*}

\begin{figure*}[hb!]
\centering     
\subfigure{\label{exp:alpha-compas}\includegraphics[width=0.7\linewidth]{figs/fig3_Legend.png}}
\subfigure[Remove NA]{\label{}\includegraphics[width=0.45\linewidth]{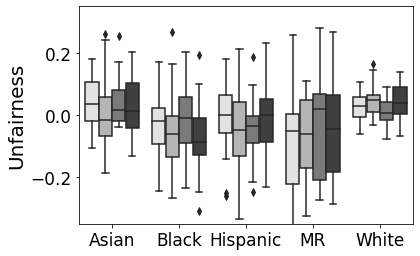}}
\subfigure[Simple Imp.]{\label{}\includegraphics[width=0.45\linewidth]{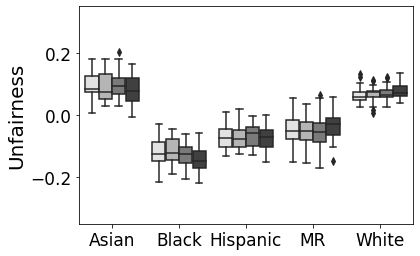}}
\subfigure[Multiple Imp.]{\includegraphics[width=0.45\linewidth]{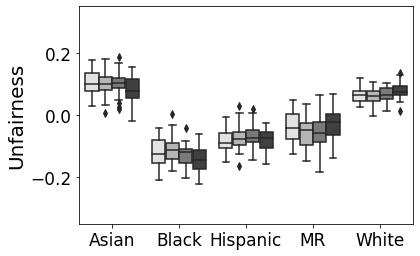}}
\subfigure[KNN Imp.]{\label{}\includegraphics[width=0.45\linewidth]{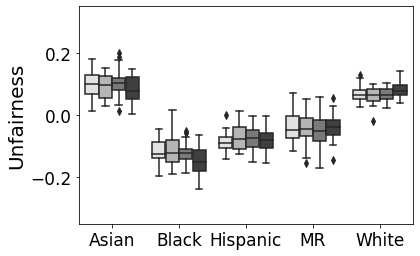}}
\caption{Impact of Imputation- Equalized odds}\label{exp:bars_eo}
\end{figure*}

\begin{figure*}[hb!]
\centering 
\subfigure{\label{exp:alpha-compas}\includegraphics[width=0.7\linewidth]{figs/fig4_Legend.pdf}}
\subfigure[Statistical Parity]{\label{exp:alpha-compas}\includegraphics[width=0.49\linewidth]{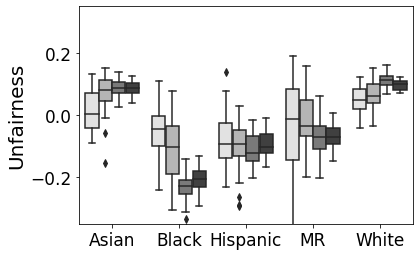}}
\subfigure[Predictive Equality]{\label{exp:nested-compas}\includegraphics[width=0.49\linewidth]{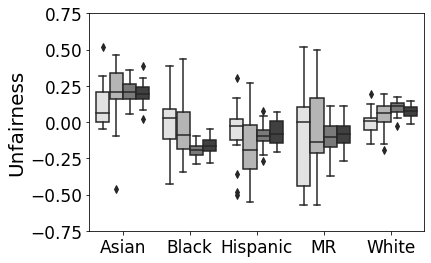}}
\subfigure[Equal Opportunity]{\label{exp:append-compas}\includegraphics[width=0.49\linewidth]{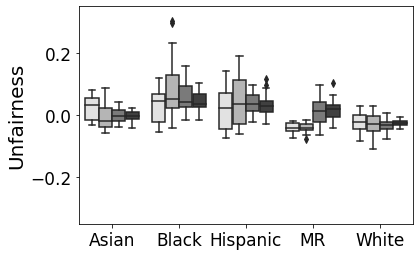}}
\subfigure[Equalized Odds]{\label{exp:alpha-adult}\includegraphics[width=0.49\linewidth]{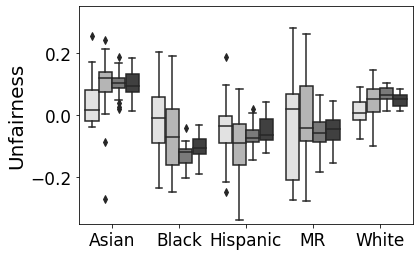}}
\subfigure[Testing Accuracy]{\label{exp:alpha-adult}\includegraphics[width=0.49\linewidth]{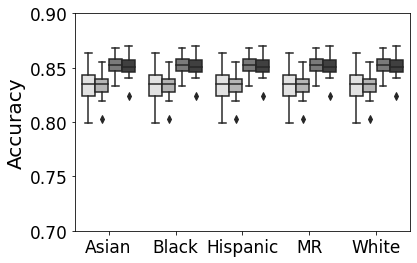}}
\caption{Remove NA vs Imputation (RF)}\label{exp:bars_sp}
\end{figure*}

\begin{figure*}[hb!]
\centering  
\subfigure{\label{exp:alpha-compas}\includegraphics[width=0.8\linewidth]{figs/fig5_Legend.png}}
\subfigure[Statistical Parity]{\label{exp:alpha-compas}\includegraphics[width=0.49\linewidth]{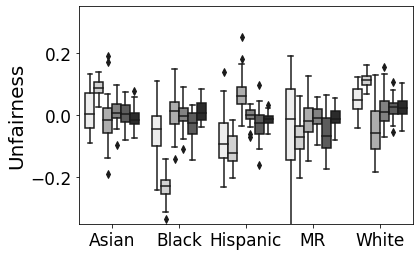}}
\subfigure[Predictive Equality]{\label{exp:nested-compas}\includegraphics[width=0.49\linewidth]{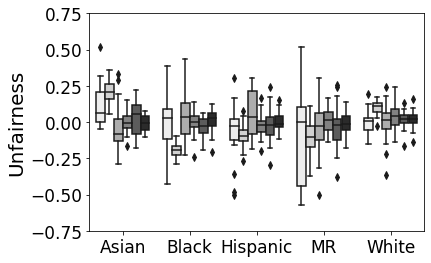}}
\subfigure[Equal Opportunity]{\label{exp:append-compas}\includegraphics[width=0.49\linewidth]{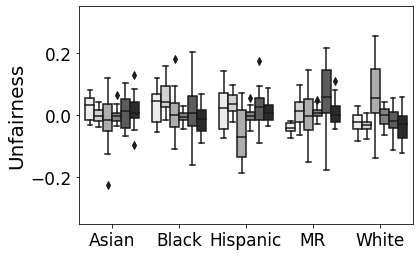}}
\subfigure[Equalized Odds]{\label{exp:alpha-adult}\includegraphics[width=0.49\linewidth]{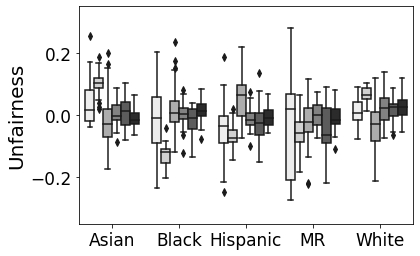}}
\subfigure[Testing Accuracy]{\label{exp:alpha-adult}\includegraphics[width=0.49\linewidth]{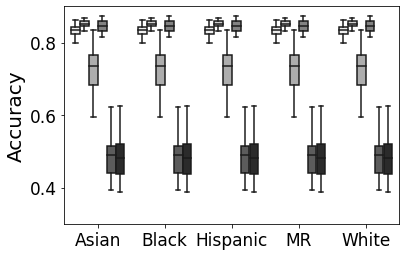}}
\caption{Fairness and Performance Evaluation of Different Perturbation Scenarios compared against Remove-NA and Imp.  (RF)}\label{exp:bars_sp}
\end{figure*}

\begin{figure*}[hb!]
\centering  
\subfigure{\label{exp:alpha-compas}\includegraphics[width=0.6\linewidth]{figs/fig6_Legend.png}}
\subfigure[Statistical Parity]{\label{exp:alpha-compas}\includegraphics[width=0.49\linewidth]{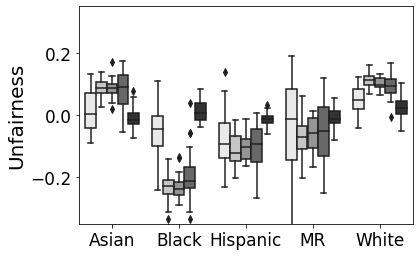}}
\subfigure[Predictive Equality]{\label{exp:nested-compas}\includegraphics[width=0.49\linewidth]{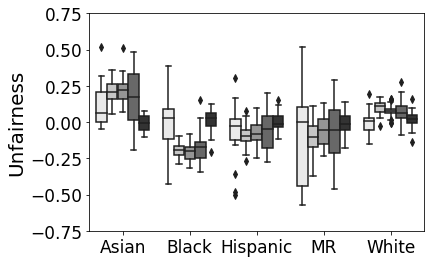}}
\subfigure[Equal Opportunity]{\label{exp:append-compas}\includegraphics[width=0.49\linewidth]{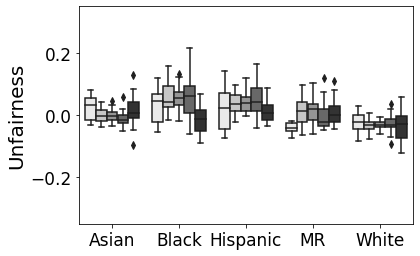}}
\subfigure[Equalized Odds]{\label{exp:alpha-adult}\includegraphics[width=0.49\linewidth]{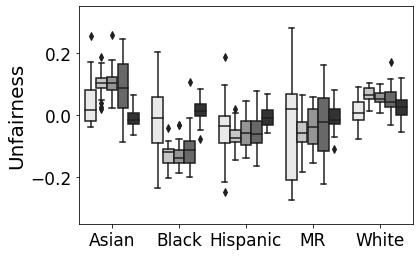}}
\subfigure[Testing Accuracy]{\label{exp:alpha-adult}\includegraphics[width=0.49\linewidth]{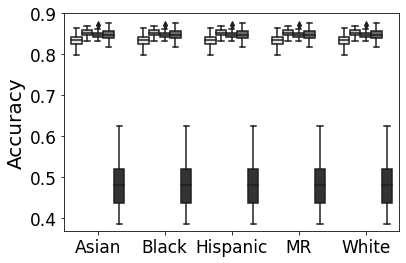}}
\caption{Fairness and Performance Evaluation of Remove-NA, Imputation Scenarios vs. the Perturbation of all attributes (RF)}\label{exp:bars_sp}
\end{figure*}





\end{document}